\documentclass[11pt]{article}

\usepackage{graphicx}
\usepackage[utf8]{inputenc}

\usepackage[english]{babel}
\usepackage{amssymb}
\usepackage{amsfonts}
\usepackage{amsmath}
\usepackage{amsthm}
\usepackage{caption}
\usepackage{mathtools}

\newtheorem{theorem}{Theorem}[section]

\newtheorem{proposition}[theorem]{Proposition}

\def\RR{\mathbb{R}}

\def\cS{{\cal S}}

\def\bp{\mbox{\boldmath $p$}}

\def\bPhi{\mbox{\boldmath $\Phi$}}

\def\ie{\textit{i.e.~}}

\def\be{\begin{equation}}
\def\ee{\end{equation}}    

\begin{document}

\title{\LARGE Causal Lie products of free fields and the \\
emergence of quantum field theory}
\author{Detlev Buchholz${}^a$, Roberto Longo${}^b$ and
  Karl-Henning Rehren${}^c$ \\[20pt]
\small  
${}^a$ Mathematisches Institut, Universit\"at G\"ottingen \\
\small Bunsenstr.\ 3-5, 37073 G\"ottingen, Germany \\[5pt] 
\small
${}^b$ 
Dipartimento di Matematica,
Università di Roma Tor Vergata  \\
\small Via della Ricerca Scientifica 1, 00133 Roma, Italy \\[5pt] 
\small
${}^c$ Institut für Theoretische Physik, Universit\"at G\"ottingen, \\
\small Friedrich-Hund-Platz 1, 37077 G\"ottingen, Germany \\[5pt] 
}
\date{}

\maketitle

\noindent \textbf{Abstract.}   
All causal Lie products of solutions of the
Klein-Gordon equation and the wave equation in Minkowski space
are determined. The results shed light on the
origin of the algebraic structures underlying quantum field theory.



\section{Introduction}
\noindent In this note we analyze the structure of
causal Lie products (brackets)
of solutions of the Klein-Gordon equation
in Minkowski space with mass $m \geq 0$. This problem was
studied in the Wightman framework of quantum field
theory by several people: for massive
free fields by R.~Jost \cite{Jo}, B.~Schroer \cite{Sch},
P.G.~Federbush, K.A.~Johnson~\cite{FeJo}, and for massless fields by
K.~Pohlmeyer~\cite{Po}. It revealed the fact
that free fields must have c-number commutators.
However, these results depend crucially on the
assumptions of some underlying Hilbert space structure and
spectral properties of a Hamiltonian (the existence of
a vacuum sector). It is the aim of
the present note to establish the 
properties of free fields in a more general framework,
shedding some light on the origin of the algebraic structures 
underlying quantum field theory. 

\section{Framework and results}
\noindent  Let $\cS(\RR^d)$ 
be the Schwartz space of real scalar test functions on
Minkowski space $\RR^d$, $d > 2$, with its standard metric. We consider
a Lie algebra $\bPhi$ over $\RR$
that is generated by symbols $\phi(f)$ which are real linear with
regard to $f \in \cS(\RR^d)$. Thus we have for their Lie products the
standard relations (anti-symmetry and Jacobi identity),
$f_1,f_2,f_3 \in \cS(\RR^d)$, 
\begin{align}
& \hspace*{-1.5mm}  [\phi(f_1), \phi(f_2)] = -  [\phi(f_2), \phi(f_1)]  \\
& \hspace*{-1.5mm} [[\phi(f_1), \phi(f_2)],\phi(f_3)] + 
  [[\phi(f_3), \phi(f_1)],\phi(f_2)] +
  [[\phi(f_2), \phi(f_3)],\phi(f_1)] = 0 \, . \nonumber
\end{align}  
In addition, we assume that $\phi$ is a solution of the Klein-Gordon
equation (including the wave equation for mass $m=0$) and that
its Lie products comply with the condition of Einstein causality
(locality). Thus we have the additional relations
for $f_1, f_2 \in \cS(\RR^d)$, 
$K \coloneqq \square + m^2$ being the Klein-Gordon operator
and $\perp$ denoting spacelike separation,
\begin{align}
  & \phi(Kf_1) = 0 \, ,  \\
  &  [\phi(f_1), \phi(f_2)] = 0 \quad \text{if} \quad 
    \text{supp} \, f_1 \perp  \text{supp} \, f_2 \, . \nonumber 
\end{align}     

\medskip
We assume that there exist (multi)linear forms on $\bPhi$ which are
tempered distributions with regard to the underlying test functions.
We also assume that the collection of all such forms on $\bPhi$ is
faithful, \ie the intersection of their kernels is trivial.
(Frobenius Lie algebras are examples with this property.) Given any
such form $l$, we want to show that
\be \label{e.3}
l([[\phi(f_1), \phi(f_2)], \phi(f_3)]) = 0 \, ,
\quad f_1,f_2,f_3 \in \cS(\RR^d) \, .
\ee
Since the collection of all functionals is faithful, this implies that 
the Lie products $[\phi(f_1), \phi(f_2)]$ lie in the center of~$\bPhi$. 

\medskip
We begin by noting that by the Schwartz kernel (nuclear)
theorem, the distribution \eqref{e.3} extends by continuity
in $f_1 \otimes f_2 \otimes f_3$ to arbitrary test functions 
$f \in \cS(\RR^{3d})$. Thus it is sufficient to consider for any   
$h \in \cS(\RR^d)$ the distributions     
\be
x,y \mapsto l_h([[\phi(x),\phi(y)], \phi(0)])
\coloneqq \int \! dz \, h(z) \, l([[\phi(x +z),\phi(y + z)], \phi(z)]) \, ,
\ee
from which the original $l$ can be recovered. 
Because of locality, $l_h$ vanishes for $(x-y)^2 < 0$ and, by the Jacobi
identity, this also obtains if, both, $x^2 < 0$ and $y^2 < 0$.
Picking any fixed spacelike $y$, it follows that the  distribution
vanishes if $x^2 < 0$ or $(x - y)^2 < 0$. Thus it vanishes with regard 
to $x$ in some open 
time slice. Since it is a solution in $x$ of the Klein-Gordon equation,
it vanishes for all $x$. 

\medskip
With this information we pick now any $x$. According to the preceding step,
the distribution then vanishes in $y$ for $y^2 < 0$ and, again 
by locality, for $(x - y)^2 < 0$. If $x$ is spacelike, 
the preceding step implies that the distribution vanishes for all $y$.

\medskip 
If $x$ is positive timelike, the distribution vanishes for $y$ in the
interior of the complement
of $(V_+ + x) \cup V_- \cup D$, where $V_\pm$ are the
forward and backward lightcones, respectively,
and $D$ is the double cone
fixed by the timelike line segment from $0$ to $x$. We pick now any
timelike line segment $L$ in this complement which does not touch $D$ and 
the boundaries of $(V_+ + x)$ and $V_-$, cf.\ Fig.~1. Since $L$
is localized in the interior of this complement, the
solution of the Klein-Gordon equation with regard to $y$ vanishes
in a timelike tube and hence in
the double cone fixed by it; this is a
consequence of standard theorems
on solutions of hyperbolic differential equations with
constant coefficients \cite{CoHi} or of Borchers'
double cone theorem~\cite{Bo}. Performing this step for
all such line segments, we find that the distribution vanishes
for all $y$ with support in the interior of the complement of
$(V_+ + x) \cup V_-$. Making again use of the fact that
the distribution is a solution of the Klein-Gordon
equation in $y$, this implies that it vanishes for all $y$.
A similar argument applies if $x$ is negative timelike. 

\begin{figure}[h]
\centering
\includegraphics[width=0.35\textwidth]{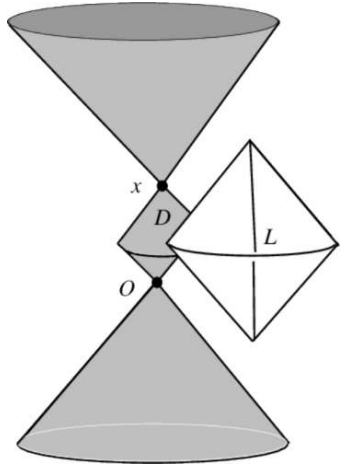}
\caption*{\small
  Fig.\ $\! \! 1$ Initial support of the Lie product (gray), which
  is further restricted by~the exis\-tence of timelike line segments $L$
  in its complement and the Klein-Gordon equation.} 
\end{figure}

We finally discuss the case that $x$ is positive lightlike. 
As in the preceding step, the distribution 
vanishes with regard to $y$
in the interior of the
complement of the region $(V_+ + x) \cup V_- \cup D$,
where $D$ now denotes the degenerate double cone consisting of
the lightlike line segment connecting $0$ and $x$. At this
point it matters that the dimension $d$ of spacetime is larger
than~$2$. If $d=2$ and the
field is massless, that complement cannot be enlarged.
This follows from the observation that in the  
quantum field theoretic example of local chiral fields their   
normal ordered products still satisfy the wave equation, but their double
commutators do not vanish. Yet if
$d > 2$, there exist timelike line segments $L$ in the interior of the above
complement that cross the characteristic hyperplane
defined by the lightlike line segment at an arbitrarily small distance from it.
The corresponding double cones fixed by $L$ then intersect the
lightlike segment in some open interval. Hence the distribution
vanishes again in $y$ in some time slice and consequently
vanishes everywhere. A similar argument works if
$x$ is negative lightlike. We summarize these results in a
first proposition. 

\begin{proposition}
  Let $\bPhi$ be a Lie algebra with properties given above. Then
  the Lie product of any two elements lies in the center of this 
  algebra. 
\end{proposition}  
It remains to determine the possible form of the
Lie products. To this
end we consider the distribution,
canonically extended to complex test functions~$h$,   
\be
x \mapsto l_h([\phi(x), \phi(0)]) \coloneqq 
\int \! dz \, h(z) \, l([\phi(x +z),\phi(z)]) \, .
\ee
It is a  solution of the Klein-Gordon equation which vanishes
because of locality if $x^2 < 0$. So its Cauchy data at time $0$ are
tempered distributions on space which are localized
at the origin. Hence they are finite sums of derivatives
of the Dirac measure, supported at $0$. It follows that
\be \label{e6}
l_h([\phi(x), \phi(0)]) = \int \! dp \, P_h(p) \, \varepsilon(p_0)
\delta(p^2 - m^2) \, e^{ipx} \, ,
\ee
where $P_h$ is a polynomial in the zero and spatial components of
$p$ of the form $p \mapsto P_h(p) = p_0 \, Q_h(\bp) + R_h(\bp)$;
it depends in a complex linear and tempered 
manner on $h$. In order to determine
the dependence of $P_h$ on $h$,
we exploit the fact that the second field in
this distribution also satisfies the Klein-Gordon equation
and proceed to 
\be
x \mapsto l_h([\phi(x), \phi(0)]) =
\int \! dz \, h(z-x) \, l([\phi(z),\phi(z - x)]) \, . 
\ee
Applying the Klein-Gordon operator, we get 
\begin{align}
& 0 = \int \! dz \, (\square_x h(z - x) + 2  \, \partial_{x \mu} h(z -x) \,
  \partial_x^\mu) \, l([\phi(z),\phi(z - x)]) \\
  & =  - \int \! dz \,  (\square h)(z) \, l([\phi(x + z),\phi(z)])  
  - 2 \, \partial_x^\mu \int \! dz \, (\partial_\mu h)(z)
  \, l([\phi(x + z),\phi(z)]) \nonumber \\
  & = - 2 \, \partial_x^\mu \, l_{\, \partial_\mu h}([\phi(x), \phi(0)])
  - l_{\, \square h}([\phi(x), \phi(0)]) \, . \nonumber
  \nonumber
\end{align}
Making use of the preceding result \eqref{e6} on the structure of $l_h$, 
we obtain for the resulting polynomials the equality on the mass shell
$p^2 = m^2$
\be
2i p^\mu P_{\, \partial_\mu h}(p) + P_{\, \square h}(p) =
P_{(2i p^\mu \partial_\mu + \square)h}(p) = 0  \, .
\ee
Thus we have for arbitrary test functions $h$
\be
P_h(p) = \int \! dq \, (A(p) \delta(q) + B(p) \delta(q^2 + 2qp))
\widetilde{h}(q) \, ,
\ee
where the tilde $\tilde{ \ \ }$ denotes Fourier transforms.
Since this expression is to be a polynomial in the components of $p$ for
arbitrary $h$, it follows 
that $B=0$ and that $A$ is a polynomial. Hence
$p \mapsto P_h(p) = \widetilde{h}(0) A(p)$, where $A$ is of the special 
form given above. Summarizing these
observations, we have arrived at our second proposition.
\begin{proposition} \label{p2}
  Let $\bPhi$ be a Lie algebra with properties given above. Then
  one has for any tempered functional $l$ the equality in the
  sense of distributions
  \be \label{e11}
  l([\phi(x), \phi(y)]) =
  \int \! dp \, A_l(p) \, \varepsilon(p_0)
\delta(p^2 - m^2) \, e^{ip(x-y)} \, ,
\ee
where $A_l$ is some even (as a consequence of the
antisymmetry of the Lie product) polynomial.
\end{proposition}  

In view of the preceding results one can fix the central elements of
the Lie algebra $\bPhi$. This is accomplished by first extending the 
algebra by complex multiples of an identity $1$ and then taking the quotient
with regard to the ideal generated by
\be
   [\phi(f), \phi(g)] - l([\phi(f), \phi(g)]) \, 1 \, , \quad 
   f,g \in \cS(\RR^d) \, .
\ee
In this manner, the Lie products
are identified with multiples of the identity
with the concrete numerical factors, given by 
Proposition \ref{p2}.
Let us emphasize that the polynomial $A_l$ in this proposition 
depends on the choice of the functional~$l$. 
Its specific form is not
encoded in the Lie algebra $\bPhi$. As a matter of fact, every even
polynomial $A$ defines by equation~~\eqref{e11} some admissible functional
on~$\bPhi$ and thereby some ``primary'' non-commutative Lie algebra
whose center consists of multiples of $1$. 

\section{Conclusions}
Starting from the assumption that the action induced by free fields on
each other can be described by a Lie algebra, in analogy to the Poisson
brackets in classical physics, we have determined the possible
realizations of these algebras which are compatible with Einstein
causality. It turned out that the Lie products of the fields are
elements of the center of the algebra with very specific
properties: all functionals on the Lie products 
are invariant under simultaneous translations of the fields and
they are finitely covariant with regard to Lorentz transformations, \ie
they transform like components of a tensor field \cite{BrEpGl}.
Any such functional determines a primary Lie algebra with c-number
commutation relations.

\medskip 
The best known examples which are compatible with these results are
the scalar free fields $\phi_0$ in quantum field theory, where the
polynomial $A_0$ appearing in the Lie product is some positive constant,
depending on the dimension $d$. In that case the Lie-algebra 
can consistently be extended to a non-commutative *-algebra
which is faithfully represented on Fock space and where the
Lie product is given by the commutator of the field operators. 

\medskip 
It turns out that any 
other ``primary'' Lie algebra for a given
even polynomial $A$ is obtained from $\phi_0$ by a real linear map 
\be
\phi_0(f) \mapsto \phi_0(B f) \, , \quad f \in \cS(\RR^d) \, .
\ee
Here $B$ is a 
multiplication operator in momentum space.
The underlying function $p \mapsto B(p)$
has the following properties: it is 
even with regard to momentum, its square equals $A$, 
and it is continuous almost
everywhere on~$\RR^d$, cf.\ the appendix. The corresponding fields are 
faithfully represented on Fock space as well.
Since $\phi_0(B f)^* \supset \phi_0(\overline{B f})$, 
these operators are not symmetric, however,
unless $B$ is real in momentum space. This can only happen 
if $A$ is positive and hence the value of the commutator 
\eqref{e11} is purely imaginary. In the general case, the fields and
also their adjoints have local commutators on Fock space, but
they are not relatively local. Alternatively, these cases
can be realized by components of tensor fields on indefinite
Fock spaces. 

\medskip
So we conclude that the appearance of the familiar algebraic 
structures of quantum field theory 
can be traced back in the case of free fields to
Einstein causality. Results pointing into a similar direction 
were also established in~\cite{BuFr}, but the existence of a
$*$-operation (of a complex structure)
was assumed there from the outset. As we have
seen here, this feature is a consequence of the representation
theory of the specific Lie algebras, which emerge from our 
assumptions. They have faithful representations by Hilbert space 
operators as a consequence of Einstein causality.

\section*{Acknowledgment}
\noindent DB  is grateful
to Dorothea Bahns and the Mathematics Institute
of the University of G\"ottingen for their continuing hospitality. 
RL thanks the Alexander von Humboldt Foundation for
supporting his visit of the University of G\"ottingen,
which made this collaboration possible. He also acknowledges an ERC
grant and the MIUR Excellence Department Project awarded to the
Department of Mathematics, University of
Rome Tor Vergata, CUP E83C18000100006.

\section*{Data availability}
Data sharing is not applicable to this article as no new data were
created or analyzed in this study.

\begin{appendix}
\section*{Appendix}
For completeness, we give here a proof of the elementary fact, used
in the main text, that any complex polynomial $A$ on $\RR^d$
(which is not necessarily of the special type considered here)
has a square root $B$ which is
continuous almost everywhere.  We proceed from
the principal square root $z \mapsto \sqrt{z}$ on the complex
plane, which is defined for $- \pi < \arg z \leq \pi$ and vanishes
at $0$. It is continuous along the real axis. In the
complement of the (closed) negative axis it is analytic
and it is discontinuous across the cut at the negative reals. 

\medskip 
With this
choice, we put $p \mapsto B(p) \coloneqq \sqrt{A(p)}$. If $A$
is symmetric in $p$, as in the present investigation,
it is clear that $B$ is symmetric as well since 
the principal square root does not depend on $p$. Discontinuities 
of $B$ can only appear at points where the values of $A$ cross the
cut at the negative reals from imaginary directions. Defining the
set
\be
S \coloneqq \{ p \in \RR^d : A(p) \leq 0 \} \, ,
\ee
there are the following possibilities: (i) \ $S = \emptyset$. Then
$B$ is continuous (even real analytic). (ii) \ $S = \RR^d$. Then
$A$ is real, hence $B$ is continuous.
\mbox{(iii) \ $S \subset \RR^d$} is a proper subset. If the
imaginary part $\, \text{Im} A$ of $A$ vanishes on $\RR^d$, then $A$
is real again and $B$ is continuous. If $\, \text{Im} A$ varies on
$\RR^d$, then, being a polynomial, it can vanish only on sets of
zero Lebesgue measure, cf.~\cite{Ca}. Thus $S$, being a closed
subset, has zero measure as well. Only in the latter case it
can happen that $B$ is discontinuous on
such negligible sets of~$\RR^d$. 

\end{appendix}

\end{document}